\documentclass[prl,aps,amsfonts,amssymb,draft,floats,
twocolumn]{revtex4}
\usepackage{epsf}

\begin{document}

\title{Symmetry of boundary conditions of the Dirac equation for
electrons in carbon nanotubes}

\author{Edward McCann and Vladimir I. Fal'ko}

\affiliation{Department of Physics, Lancaster University,
Lancaster, LA1 4YB, United Kingdom}

\begin{abstract}
We consider the effective mass model of spinless electrons in single wall
carbon nanotubes that is equivalent to the Dirac equation for massless fermions.
Within this framework we derive all possible energy independent hard wall boundary
conditions that are applicable to metallic tubes.
The boundary conditions are classified in terms of their symmetry properties
and we demonstrate that the use of different boundary conditions will result
in varying degrees of valley degeneracy breaking of the single particle energy
spectrum.
\end{abstract}

\maketitle

%%%%%%%%%%%%%%%%%%%%%%%%%%%%%%%%%%%%%%%%%%%%%%%%%%%%%%%%%%%%%%%%%%%%
%%%%%%%%%%%%%%%%%%%%%%%%%%%%%%%%%%%%%%%%%%%%%%%%%%%%%%%%%%%%%%%%%%%%
\section{Introduction}
%%%%%%%%%%%%%%%%%%%%%%%%%%%%%%%%%%%%%%%%%%%%%%%%%%%%%%%%%%%%%%%%%%%%
%%%%%%%%%%%%%%%%%%%%%%%%%%%%%%%%%%%%%%%%%%%%%%%%%%%%%%%%%%%%%%%%%%%%

Carbon nanotubes are the subject of intense research, motivated by the
desire to use their unique physical and electronic properties in the
development of nanoscale electrical devices\cite{saito98,dek99}. The
electronic properties of nanotubes follow from the band structure of
graphene - a two-dimensional (2D) sheet of graphite - which is a semi-metal,
having a vanishing energy gap at the six corners, K-points, of the hexagonal
first Brillouin zone. A single-wall nanotube may be thought of as a graphene
sheet rolled up to form a nanometre-diameter cylinder. Periodicity around
the circumference results in quantized transverse wavevectors leading to
metallic or semiconducting behaviour depending on whether the K-point
wavevector ${\bf K}$ is an allowed wavevector.

While the energy spectrum of an infinitely long tube will be continuous, a
finite tube should possess discrete energy levels corresponding to standing
waves typical of a confined quantum particle. Evidence of discrete levels
was seen in transport measurements \cite{bock97,tans97} a few years ago,
followed by the direct observation of sinusoidal standing wave patterns by
scanning tunneling microscopy\cite{ven99,lem01}. The measured wavelength of
the standing waves $\lambda \sim 0.75$nm, about three times larger than the
lattice constant $a\approx 0.25$nm, corresponded to wavevectors near the
K-point ${\bf K}$.
More recently, Coulomb blockade measurements on carbon nanotube quantum
dots \cite{lia02,bui02,cob02} have found evidence for fourfold periodicity of
the spectra that is in agreement with expectations based on spin and K-point
degeneracy, although the experiments appeared to show varying degrees of
degeneracy breaking.
A number of authors \cite{rubio99,roche99,wu00,y+a01,jiang02}
have modelled finite-length nanotubes in order to describe the atomic scale
variation of standing waves patterns and the opening of an energy gap
displaying an oscillating dependence on the tube length.
Rather than concentrating on one particular model of a boundary,
we aim to describe all possible energy independent hard wall
boundary conditions for metallic single wall nanotubes. We will classify
the boundary conditions in terms of their symmetry properties and show how
different boundary conditions produce varying degrees of K-point
degeneracy breaking.

In the scanning tunneling microscopy measurements of Ref.~\onlinecite{lem01} an
additional slow spatial modulation of the standing waves was observed. It
was interpreted as being a beating envelope function with wavevector ${\bf q}$,
$\left| {\bf q}\right| \ll \left| {\bf K}\right| $, resulting from
the interference of left and right moving waves with slightly different total
wavevectors ${\bf K}\pm {\bf q}$. Theoretically, the effective mass
model \cite{d+m84,a+a93,k+m97,mceuen99} provides a reliable analytical
description of the electronic structure near the K point where the total
wavevector is ${\bf k}={\bf K}+{\bf q}$ and the dispersion relation is linear
$E= sv\left| {\bf q}\right| $, $v$ is the Fermi velocity and $s = \pm 1$
for the conduction and valence band, respectively. For spinless
electrons, the envelope wavefunction $\Psi \left( {\bf q},{\bf r}\right)$ has four
components corresponding to two inequivalent atomic sites in the hexagonal
graphite lattice (``A'' and ``B'') and to two inequivalent K-points in the
hexagonal first Brillouin zone.
The resulting eigenvalue equation for $\Psi$ is the massless Dirac equation,
\begin{eqnarray}
-iv{\bf \alpha .\nabla }\Psi  &=& E\Psi ; \hspace{0.4in}
\alpha = \left( 
\begin{array}{cc}
{\bf \sigma } & 0 \\ 
0 & -{\bf \sigma }
\end{array}
\right) ; \\ \label{diracintro} 
{\bf \sigma } &=& e^{i\eta \sigma _{z}/2}\left( \sigma _{x}{\bf \hat{\imath}}
+\sigma _{y}{\bf \hat{\jmath}}\right) e^{-i\eta \sigma _{z}/2} \nonumber 
\end{eqnarray}
where the role of spin (``pseudo-spin'') is assumed by the relative
amplitudes on the A and B atomic sites: ${\bf \sigma }$ is a vector in the
$(x,y)$ plane rotated by the chiral angle $\eta$ of the tube.
Also, $v=\left( \sqrt{3}/2\right) a\gamma$ is the Fermi velocity,
$a$ is the lattice constant of graphite and $\gamma$
is the nearest neighbour transfer integral.

In this paper we consider the effective boundary conditions for the envelope
function $\Psi$ in a finite size carbon nanotube.
Since the effective mass model for $\Psi$
corresponds to the Dirac equation, we begin by deriving all possible energy
independent hard wall boundary conditions for the Dirac equation. We write
them in terms of a small number of arbitrary parameters, mixing angles, that
describe mixing between boundary conditions with different discrete
symmetries. Then, in order to illustrate the meaning of the general boundary
conditions, we evaluate the resulting energy level spectra for
non-interacting electrons in finite-length metallic nanotubes and the
corresponding standing wave envelope functions.
To anticipate a little, we find that energy independent hard wall boundary
conditions for the Dirac equation may be expressed in general terms as 
\begin{equation}
\Psi =M\Psi ;\hspace{0.4in}M^{2}=1;\hspace{0.4in}\left\{ {\bf n}_{{\bf B}}
{\bf .\alpha },M\right\} =0,  \label{bcintro}
\end{equation}
where $M$ is an Hermitian, unitary $4\times 4$ matrix $M^{2}=1$ with the
constraint that it anticommutes with the operator
${\bf n}_{{\bf B}}{\bf .\alpha }$, proportional to the component of the current
operator normal to the interface, ${\bf n}_{{\bf B}}$ is the unit vector normal
to the interface.
As explained in the Appendix, we find four possible linear combinations of
matrices satisfying these constraints on $M$, which, assuming
${\bf n}_{{\bf B}}$ is a vector confined to the $\left( x,y\right) $ plane,
may be written in terms of a small number of arbitrary parameters:
\begin{eqnarray}
M_{1} &=&\cos \Lambda \left( I_{\Pi }\otimes {\bf n}_{1}{\bf .\sigma }
\right) +\sin \Lambda \left( \Pi _{z}\otimes {\bf n}_{{\bf 2}}{\bf .\sigma }
\right) ,  \label{m1} \\
M_{2} &=&\cos \Upsilon \left( {\bf \nu }_{{\bf 1}}{\bf .\Pi }\otimes
I_{\sigma }\right) +\sin \Upsilon \left( {\bf \nu }_{{\bf 2}}{\bf .\Pi }
\otimes {\bf n}_{{\bf B}}{\bf .\sigma }\right) ,  \label{m2} \\
M_{3} &=&\cos \Omega \left( {\bf \nu }_{{\bf 2}}{\bf .\Pi }\otimes {\bf n}_{
{\bf B}}{\bf .\sigma }\right) +\sin \Omega \left( I_{\Pi }\otimes {\bf n}_{
{\bf 1}}{\bf .\sigma }\right) ,  \label{m3} \\
M_{4} &=&\cos \Theta \left( {\bf \nu }_{{\bf 1}}{\bf .\Pi }\otimes I_{\sigma
}\right) +\sin \Theta \left( \Pi _{z}\otimes {\bf n}_{{\bf 2}}{\bf .\sigma }
\right) ,  \label{m4}
\end{eqnarray}
where the angles $\Lambda $,$\Upsilon $,$\Theta $ and $\Omega $ are
arbitrary, ${\bf n}_{{\bf 1}}$ and ${\bf n}_{{\bf 2}}$ are three-dimensional
space-like vectors satisfying the contraints ${\bf n}_{{\bf 1}}{\bf .n}_{
{\bf B}}={\bf n}_{{\bf 2}}{\bf .n}_{{\bf B}}={\bf n}_{{\bf 1}}.{\bf n}_{{\bf 
2}}=0$, and ${\bf \nu }_{{\bf 1}}$ and ${\bf \nu }_{{\bf 2}}$ are
two-dimensional (confined to the $\left( x,y\right) $ plane) space-like
vectors satisfying the constraint ${\bf \nu }_{{\bf 1}}{\bf .\nu }_{{\bf 2}
}=0$. Here we have adopted a matrix direct product notation to highlight the
separate K-point space and AB space structure, using the notation $\left\{
\sigma _{x},\sigma _{y},\sigma _{z},I_{\sigma }\right\} $ for $2\times 2$
Pauli matrices and the unit matrix that operate within a block (`AB space')
and $\left\{ \Pi _{x},\Pi _{y},\Pi _{z},I_{\Pi }\right\} $ for $2\times 2$
Pauli matrices and the unit matrix that operate in K-point space. For
example, the operator $\alpha $ may be written as a direct product $\alpha
=\Pi _{z}\otimes {\bf \sigma }$. Note that the boundary conditions of the
Hadron bag model\cite{chodos74}, in which elementary particles are confined
by a scalar (mass) term at the boundary, are described by
$M = {\bf \nu }_{{\bf 2}}{\bf .\Pi }\otimes {\bf n}_{{\bf B}}{\bf .\sigma }$.
Berry and Mondragon \cite{berry87} considered
``neutrino billiards'' with a two component Dirac equation confined by a
term proportional to $\sigma _{z}$, corresponding to
$M = I_{\Pi }\otimes {\bf n}_{1}{\bf .\sigma }$ or 
$M= \Pi _{z}\otimes {\bf n}_{{\bf 2}}{\bf .\sigma }$ with either
${\bf n_1}$ or ${\bf n_2}$ lying in the $\left( x,y\right)$ plane.

There are two non-equivalent K-points that we label as $K$ and $\widetilde{K}$.
The Dirac equation is diagonal in K-point space, so that, in the absence of
boundary conditions, there are two right moving ($\Psi _{K}^{(R)}$ and $\Psi
_{\widetilde{K}}^{(R)}$) and two left moving ($\Psi _{K}^{(L)}$ and $\Psi _{
\widetilde{K}}^{(L)}$) plane wave solutions near the Fermi surface
of a metallic tube.
The solutions $\Psi_{K}^{(R)}$\ and $\Psi_{\widetilde{K}}^{(L)}$\ are
eigenvectors of the pseudo-spin component along the tube axis
${\bf \Sigma .n_{B}}$ with eigenvalue $+s$, whereas the solutions
$\Psi_{\widetilde{K}}^{(R)}$ and $\Psi_{K}^{(L)}$ have eigenvalue $-s$,
where $s = \pm 1$ for the conduction and valence band, respectively. Also, the
solutions $\Psi_{K}^{(R)}$\ and $\Psi_{K}^{(L)}$\ are eigenvectors of
pseudo-helicity $-i{\bf \Sigma .\nabla }/|{\bf q}|$\ with eigenvalue $+s$,
whereas the solutions $\Psi_{\widetilde{K}}^{(R)}$ and
$\Psi_{\widetilde{K}}^{(L)}$ have eigenvalue $-s$.
There are different ways of combining the right and
left moving waves in order to create standing waves. The first possibility
is that waves at the same K-point combine, namely $\Psi _{K}^{(R)}$\ and $
\Psi _{K}^{(L)}$ form a standing wave with helicity eigenvalue $+s$, and $
\Psi _{\widetilde{K}}^{(R)}$ and $\Psi _{\widetilde{K}}^{(L)}$ form a
standing wave with helicity eigenvalue $-s$. This situation is realised by
the matrix $M_1$, Eq.(\ref{m1}), because it is diagonal in K-point space. A
second possibility is that waves from opposite K-points combine, namely $
\Psi _{K}^{(R)}$\ and $\Psi _{\widetilde{K}}^{(L)}$ form a standing wave
with spin component eigenvalue $+s$, and $\Psi _{\widetilde{K}}^{(R)}$ and $
\Psi _{K}^{(L)}$ form a standing wave with spin component eigenvalue $-s$.
This situation is realised by the matrix $M_2$, Eq.(\ref{m2}), because it is
off-diagonal in K-point space. A third possibility is a combination of the
previous two, with waves scattered back at the boundary into a mixture of both
of the K-points. This situation is realised by the matrices $M_3$,
Eq.(\ref{m3}), and $M_4$, Eq.(\ref{m4}), because they have both diagonal and
off-diagonal in K-point space parts.

%%%%%%%%%%%%%%%%%%%%%%%%%%%%%%%%%%%%%%%%%%%%%%%%%%%%%%%%%%%%%%%%%%%%
%%%%%%%%%%%%%%%%%%%%%%%%%%%%%%%%%%%%%%%%%%%%%%%%%%%%%%%%%%%%%%%%%%%%
\section{Effective mass model}
%%%%%%%%%%%%%%%%%%%%%%%%%%%%%%%%%%%%%%%%%%%%%%%%%%%%%%%%%%%%%%%%%%%%
%%%%%%%%%%%%%%%%%%%%%%%%%%%%%%%%%%%%%%%%%%%%%%%%%%%%%%%%%%%%%%%%%%%%

In the effective mass model of two-dimensional graphite \cite{d+m84},
the total wavefunction $\Psi_{tot}$\ is written as a
linear combination of four components $m = \left\{ 1,2,3,4\right\}$
corresponding to two K-points $\mu = \left\{ 1,2\right\}$ and $\pi$-type
atomic orbitals $\varphi_{j}({\bf r}-{\bf R_j})$ on two non-equivalent atomic
sites $j = \left\{ A,B\right\}$ in the unit cell,
\begin{equation}
\Psi_{tot}\left( {\bf r}\right) =
\sum_{m=1}^{4}\left\{ \Phi _{m}^{(0)}\left( {\bf r}\right) - {\bf G}_{m}
\left( {\bf r}\right) .{\bf \nabla }+\ldots \right\}
\psi _{m}\left( {\bf r}\right) ,  \label{totalwf}
\end{equation}
where
\begin{eqnarray}
\Phi _{m}^{(0)}\left( {\bf r}\right) &=&\frac{1}{\sqrt{N}}\sum_{{\bf R}_{
{\bf j}}}^{N}e^{i{\bf K_{\mu }.R}_{{\bf j}}}\varphi _{j}({\bf r}-{\bf R}
_{j}),  \label{phi0} \\
{\bf G}_{m}\left( {\bf r}\right) &=&\frac{1}{\sqrt{N}}\sum_{{\bf R}_{{\bf j}
}}^{N}e^{i{\bf K_{\mu }.R}_{{\bf j}}}\varphi _{j}({\bf r}-{\bf R}_{j})({\bf r
}-{\bf R}_{j}),  \label{gm}
\end{eqnarray}
are Bloch type functions constructed from the atomic orbitals,
${\bf R}_{j}$ is the position of an atom in real space and the
summation is over the number of unit cells $N\gg 1$.
The functions $\psi_{m}\left( {\bf r}\right)$ are components of the
envelope function $\Psi\left( {\bf q},{\bf r}\right)$. Substituting
this expression for $\Psi_{tot}$\ into the Schr\"{o}dinger equation
and integrating with respect to fast degrees of freedom that vary on the
scale of the unit cell leads to the Dirac equation Eq.(\ref{diracintro})
for the envelope function $\Psi$ where the K-points are taken as
${\bf K} = \left( \pm 4 \pi / 3a , 0 \right)$ and the components of $\Psi$
are written in the order $KA$, $KB$, $\widetilde{K}B$, $\widetilde{K}A$.
The appearance of the chiral angle of the tube $\eta$ in the Dirac
equation shows that the axes of the $(x,y)$ coordinate system have been
rotated to be transverse and parallel to the tube axis.
Applying periodic boundary conditions to the wavefunction $\Psi_{tot}$,
Eq.(\ref{totalwf}), in the direction transverse to the nanotube axis
produces a condition for the envelope function $\Psi$ that leads to
metallic or semiconducting behaviour depending on whether the transverse
component of wavevector ${\bf q}$ is allowed to be zero \cite{a+a93,k+m97}.

%%%%%%%%%%%%%%%%%%%%%%%%%%%%%%%%%%%%%%%%%%%%%%%%%%%%%%%%%%%%%%%%%%%%
%%%%%%%%%%%%%%%%%%%%%%%%%%%%%%%%%%%%%%%%%%%%%%%%%%%%%%%%%%%%%%%%%%%%
\section{Boundary conditions for the effective mass model}
%%%%%%%%%%%%%%%%%%%%%%%%%%%%%%%%%%%%%%%%%%%%%%%%%%%%%%%%%%%%%%%%%%%%
%%%%%%%%%%%%%%%%%%%%%%%%%%%%%%%%%%%%%%%%%%%%%%%%%%%%%%%%%%%%%%%%%%%%

In order to obtain hard wall boundary conditions for the Dirac equation, we
place an additional confinement potential at the boundary ${\bf r}={\bf r}_{B}$,
\begin{eqnarray}
\left[ -iv{\bf \alpha .\nabla }+cv\widetilde{M}\delta
\left( {\bf r-r}_{{\bf 
B}}\right) \right] \Psi =E\Psi , 
\end{eqnarray}
where $c$ is a real constant and $\widetilde{M}$ is an arbitrary $4\times 4$
Hermitian, unitary matrix, $\widetilde{M}^{2}=1$. The orientation
of the boundary is defined by a unit vector ${\bf n}_{B}$ normal to it, and
we assume that the wavefunction is zero outside the confined region, but
non-zero inside it. Then, we integrate across an infinitesimal width of the
boundary, giving
\begin{eqnarray}
-i{\bf n}_{{\bf B}}{\bf .\alpha }\Psi =c\widetilde{M}\Psi .
\end{eqnarray}
Substituting this equation back into itself, we find the
requirements that $c^{2}\widetilde{M}^{2}=1$ (thus we set $c=1$) and
$\left\{ {\bf n}_{{\bf B}}{\bf .\alpha },\widetilde{M}\right\} =0$.
The boundary condition can be written as $\Psi =M\Psi $ where
$M=i{\bf n}_{{\bf B}}{\bf .\alpha }\widetilde{M}$, $M^{2}=1$ and
$\left\{ {\bf n}_{{\bf B}}{\bf .\alpha },M\right\} =0$,
giving the result quoted in the introduction Eq.(\ref{bcintro}).

If, in the graphite coordinate system, we define the normal to the boundary as
${\bf n}_{{\bf B}}= (\sin \eta ,\cos \eta ,0)$,
where $\eta $\ is the chiral angle of the tube,
then we may choose two mutually orthogonal 3D vectors as
${\bf n_1}=\left(\cos \eta \sin \zeta ,-\sin\eta\sin\zeta ,\cos\zeta\right)$ and
${\bf n_2}=\left( \cos \eta \cos \zeta , -\sin\eta\cos\zeta,-\sin\zeta\right)$,
and two additional orthogonal 2D vectors as
${\bf \nu_1}=\left( \cos \xi ,\sin \xi, 0\right)$
and
${\bf \nu_2}=\left( -\sin \xi ,\cos \xi ,0\right)$.
This introduces two new mixing angles, $\zeta$ and $\xi$:
the arbitrary parameters contained within the boundary conditions
describe the amount of mixing between different discrete symmetries.
First, we note that the pseudo-spin of a 2D graphite sheet does not transform
in the same way as the spin of relativistic fermions, because certain
transformations result in a swapping of the orientation of A and B atoms.
This additional operation is described by a ``pseudo-spin-flip''
operator $\rho_z = \Pi_{x} \otimes i\sigma_z$ that corresponds to a reflection
in the $(x,y)$ plane of relativistic fermions.
For example, an active rotation of the 2D graphite sheet
anticlockwise by $\pi /3$ about the perpendicular $z$ axis,
$\Psi \left( {\bf r^{\prime}} \right) = C_6 \Psi \left( {\bf r} \right)$,
is described by
$C_6 = \rho_z R(\pi /3) = \Pi_{x} \otimes\exp \left( (2\pi i/3) \sigma_z \right)$
where $R(\theta ) = I_{\Pi} \otimes \exp \left( (i\theta /2) \sigma_z \right)$
is a continuous rotation operator.

Table~1 shows a summary of the discrete symmetries of the boundary conditions
in terms of the orientation of the vectors ${\bf n_1}$, ${\bf n_2}$,
${\bf \nu_1}$ and ${\bf \nu_2}$.
In addition to $\rho_z$ we consider parity $P =  \Pi_{x} \otimes I_{\sigma}$,
corresponding to a rotation by $\pi$ about the $z$ axis
($x \rightarrow -x$ and $y \rightarrow - y$),
and charge conjugation ($C$) and time reversal symmetry ($T$) that involve the
complex conjugation operator combined with $C = -i \Pi_y \otimes \sigma_y$
and $T = I_{\Pi} \otimes \sigma_y$, respectively.
The angles $\zeta$ and $\xi$ mix terms with different symmetry with respect
to $\rho_z$: $\zeta = 0$ and $\xi = 0$ correspond to evenness with respect to
$\rho_z$ whereas $\zeta = \pi /2$ and $\xi = \pi /2$ correspond to oddness.
Since spin and/or helicity label different states at the same energy,
values of $\zeta$ and $\xi$ not equal to multiples of $\pi /2$ will
lead to broken degeneracy.
The angles $\Lambda $,$\Upsilon $,$\Theta $ and $\Omega $ mix different
symmetries with respect to combinations of $P$, $C$ and $\rho_z$.

\bigskip

\centerline{
\begin{tabular}{|c|c|c|c|c|c|c|}
\hline\hline
$M$ &  &  & $\rho_{z}$ & $P$ & $C$ & $T$ \\ \hline\hline
$I_{\Pi }\otimes {\bf n}_{{\bf 1}}{\bf .\sigma }$ &
${\bf n}_{{\bf 1}}={\bf n}_{\left( x,y\right) }$ & 
$\zeta _{u}=\frac{\pi }{2}$ &
$-1$ & $+1$ & $+1$ & $-1$  \\ \hline
& ${\bf n}_{{\bf 1}}={\bf n}_{z}$ & $\zeta _{u}=0$ &
$+1$ & $+1$ & $+1$ & $-1$  \\ \hline
$\Pi _{z}\otimes {\bf n}_{{\bf 2}}{\bf .\sigma }$ &
${\bf n}_{{\bf 2}}={\bf n}_{\left( x,y\right) }$ &
$\zeta _{u}=0$ & $+1$ & $-1$ & $-1$ & $-1$  \\ \hline
& ${\bf n}_{{\bf 2}}={\bf n}_{z}$ & $\zeta _{u}=\frac{\pi }{2}$ & $-1$ & $-1$
& $-1$ & $-1$  \\ \hline
${\bf \nu }_{{\bf 1}}{\bf .\Pi }\otimes I_{\sigma }$ &
${\bf \nu }_{{\bf 1}}{\bf =\hat{\imath}}$ & $\xi _{u}=0$ &
$+1$ & $+1$ & $+1$ & $+1$  \\ \hline
& ${\bf \nu }_{{\bf 1}}{\bf =\hat{\jmath}}$ & $\xi _{u}=\frac{\pi }{2}$ &
$-1$ & $-1$ & $+1$ & $-1$  \\ \hline
${\bf \nu }_{{\bf 2}}{\bf .\Pi }\otimes {\bf n}_{{\bf B}}{\bf .\sigma }$ &
${\bf \nu }_{{\bf 2}}{\bf =\hat{\imath}}$ & $\xi _{u}=-\frac{\pi }{2}$ & $-1$
& $+1$ & $-1$ & $-1$  \\ \hline
& ${\bf \nu }_{{\bf 2}}{\bf =\hat{\jmath}}$ & $\xi _{u}=0$ &
$+1$ & $-1$ & $-1$ & $+1$  \\ \hline\hline
\end{tabular}}

\bigskip

\centerline{Discrete symmetries of the boundary conditions}

%%%%%%%%%%%%%%%%%%%%%%%%%%%%%%%%%%%%%%%%%%%%%%%%%%%%%%%%%%%%%%%%%%%%
%%%%%%%%%%%%%%%%%%%%%%%%%%%%%%%%%%%%%%%%%%%%%%%%%%%%%%%%%%%%%%%%%%%%
\section{Single particle energy spectrum}
%%%%%%%%%%%%%%%%%%%%%%%%%%%%%%%%%%%%%%%%%%%%%%%%%%%%%%%%%%%%%%%%%%%%
%%%%%%%%%%%%%%%%%%%%%%%%%%%%%%%%%%%%%%%%%%%%%%%%%%%%%%%%%%%%%%%%%%%%

In order to illustrate the meaning of the general boundary
conditions, we calculate the form of non-interacting single particle
standing waves created by the boundary conditions and the corresponding
energy spectrum. For simplicity, we will consider only metallic nanotubes
with arbitrary chiral angle $\eta $.
We suppose that the $x$ axis is perpendicular to the tube axis and we
consider only the zero momentum transverse mode so that
$|E| < 2 \pi v /|{\bf C_h}|$ where $|{\bf C_h}|$ is the circumference.
The Dirac equation is diagonal in K-point space, so that, in the absence of
boundary conditions, there are two right moving ($\Psi _{K}^{(R)}$ and $\Psi
_{\widetilde{K}}^{(R)}$) and two left moving ($\Psi _{K}^{(L)}$ and $\Psi _{
\widetilde{K}}^{(L)}$) plane wave solutions:
\begin{eqnarray*}
\Psi _{K}^{(R)} &=&Ae^{iqy}\left(
\begin{array}{c}
1 \\ 
ise^{-i\eta } \\ 
0 \\ 
0
\end{array}
\right) ; \Psi _{K}^{(L)}=Be^{-iqy}\left( 
\begin{array}{c}
1 \\ 
-ise^{-i\eta } \\ 
0 \\ 
0
\end{array}
\right) ; \\
\Psi _{\widetilde{K}}^{(R)} &=&Ce^{iqy}\left( 
\begin{array}{c}
0 \\ 
0 \\ 
1 \\ 
-ise^{-i\eta }
\end{array}
\right) ; \Psi _{\widetilde{K}}^{(L)}=De^{-iqy}\left( 
\begin{array}{c}
0 \\ 
0 \\ 
1 \\ 
ise^{-i\eta }
\end{array}
\right) ,
\end{eqnarray*}
where $A$, $B$, $C$ and $D$ are arbitrary constants, $q$ is the wavevector
along the tube and we consider $q\geq 0$ and $E=svq$, $s=\pm 1$.
The solutions $\Psi _{K}^{(R)}$\ and $\Psi_{\widetilde{K}}^{(L)}$\ are
eigenvectors of pseudo-spin component ${\bf \Sigma .\hat{\jmath}}=
I_{\Pi }\otimes e^{i\eta \sigma _{z}/2}\sigma_{y}e^{-i\eta \sigma _{z}/2}$
with eigenvalue $+s$, whereas the solutions $\Psi _{\widetilde{K}}^{(R)}$ and
$\Psi _{K}^{(L)}$ have eigenvalue $-s$. 
Also, the solutions $\Psi _{K}^{(R)}$\ and $\Psi _{K}^{(L)}$\ are eigenvectors
of pseudo-helicity
$-i{\bf \Sigma .\nabla }/|{\bf q}|= |{\bf q}|^{-1} I_{\Pi }\otimes
e^{i\eta\sigma _{z}/2} (-i\sigma_{y}\partial_{y} ) e^{-i\eta \sigma _{z}/2}$
with eigenvalue $+s$,
whereas the solutions $\Psi _{\widetilde{K}}^{(R)}$ and
$\Psi _{\widetilde{K}}^{(L)}$ have eigenvalue $-s$.
In the following we consider each of the four linear combinations
$M_1$ to $M_4$ separately, and we consider a system with the same type
of boundary condition on the right (at $y = + L/2$) and on the left
(at $y = - L/2$).
We introduce an index $u = \{R,L\} \equiv \pm 1$ to label
the right and left hand side so that the normal to the boundary,
defined with respect to the graphite coordinate system, is
${\bf n}_{{\bf B}}=u(\sin \eta ,\cos \eta ,0)$,
and we take into account the possibility of different mixing angles,
$\Lambda_u$,$\Upsilon_u$,$\Theta_u$ and $\Omega_u$, and vectors
${\bf n_1}=\left( u\cos\eta\sin\zeta_{u},-u\sin\eta\sin\zeta_{u},
\cos\zeta_{u}\right)$,
${\bf n_2}=\left( u\cos\eta\cos\zeta_{u},-u\sin\eta\cos\zeta_{u},
-\sin\zeta_{u}\right)$,
${\bf \nu_1}=\left( \cos\xi_{u},\sin\xi_{u},0\right)$
and ${\bf \nu_2}=\left( -\sin\xi_{u},\cos\xi_{u},0\right)$.

%%%%%%%%%%%%%%%%%%%%%%%%%%%%%%%%%%%%%%%%%%%%%%%%%%%%%%%%
\subsection{M1: diagonal boundary conditions}
%%%%%%%%%%%%%%%%%%%%%%%%%%%%%%%%%%%%%%%%%%%%%%%%%%%%%%%%

With the above definitions of the mixing angles,
the boundary condition $\Psi =M_{1}\Psi$ produces the following relations
between the components of the wavefunction at the interface:
\begin{eqnarray*}
u\sin \left( \zeta _{u}+\Lambda _{u}\right) e^{-i\eta }\psi _{AK}-\left[
1+\cos \left( \zeta _{u}+\Lambda _{u}\right) \right] \psi _{BK} &=&0, \\
u\sin \left( \zeta _{u}-\Lambda _{u}\right) e^{+i\eta }\psi _{A\widetilde{K}
}-\left[ 1-\cos \left( \zeta _{u}-\Lambda _{u}\right) \right] \psi _{B
\widetilde{K}} &=&0.
\end{eqnarray*}
The equations are diagonal in K-point space so do not describe intervalley
scattering. With these boundary conditions on the right (at $y = + L/2$) and
on the left (at $y = - L/2$), standing waves at $K$ are created from combining
$\Psi_{K}^{(R)}$ and $\Psi _{K}^{(L)}$ and are labelled by helicity
$\lambda = + s$, and those at $\widetilde{K}$ are created from
$\Psi _{\widetilde{K}}^{(R)}$ and $\Psi _{\widetilde{K}}^{(L)}$ and
have label $\lambda = - s$.
We find that
\begin{eqnarray*}
B &=&\left( -1\right) ^{p_{1}}A\exp \left[ is\left( \zeta _{R}-\zeta
_{L}\right) /2+is\left( \Lambda _{R}-\Lambda _{L}\right) /2\right] , \\
D &=&\left( -1\right) ^{p_{2}}C\exp \left[ -is\left( \zeta _{R}-\zeta
_{L}\right) /2+is\left( \Lambda _{R}-\Lambda _{L}\right) /2\right] ,
\end{eqnarray*}
and the corresponding wavevectors are
\begin{eqnarray*}
q^{(\lambda =+s)} &=&-\frac{s\left(\zeta_{R}+\zeta_{L}\right)}{2L}-
\frac{s\left(\Lambda _{R}+\Lambda _{L}\right) }{2L}+\frac{\pi p_{1}}{L}, \\
q^{(\lambda =-s)} &=&+\frac{s\left(\zeta_{R}+\zeta_{L}\right)}{2L}-
\frac{s\left(\Lambda _{R}+\Lambda _{L}\right) }{2L}+\frac{\pi p_{2}}{L},
\end{eqnarray*}
where $\left\{ p_{1},p_{2}\right\} $ are integers such that $q\geq 0$.
Using $E=svq$ shows that the mixing angles $\zeta _{R}$ and $\zeta _{L}$ break
K-point degeneracy whereas
$\Lambda_{R}$ and $\Lambda _{L}$ break electron-hole symmetry.

%%%%%%%%%%%%%%%%%%%%%%%%%%%%%%%%%%%%%%%%%%%%%%%%%%%%%%%%
\subsection{M2: off-diagonal boundary conditions}
%%%%%%%%%%%%%%%%%%%%%%%%%%%%%%%%%%%%%%%%%%%%%%%%%%%%%%%%

The boundary condition $\Psi =M_{2}\Psi $ is equivalent to the following
relations between the components of the enveloped wavefunction at the interface:
\begin{eqnarray*}
\psi _{AK}+u\sin \Upsilon _{u}e^{+i\eta -i\xi _{u}}\psi _{A\widetilde{K}
}-\cos \Upsilon _{u}e^{-i\xi _{u}}\psi _{B\widetilde{K}} &=&0, \\
\psi _{BK}-u\sin \Upsilon _{u}e^{-i\eta -i\xi _{u}}\psi _{B\widetilde{K}
}-\cos \Upsilon _{u}e^{-i\xi _{u}}\psi _{A\widetilde{K}} &=&0.
\end{eqnarray*}
The equations are off-diagonal in K space so describe intervalley scattering.
Standing waves are created from combining $\Psi _{K}^{(R)}$ and
$\Psi _{\widetilde{K}}^{(L)}$, with spin eigenvalue $\Sigma =+s$,
and $\Psi _{\widetilde{K}}^{(R)}$ and $\Psi_{K}^{(L)}$,
with spin eigenvalue $\Sigma =-s$.
We find that
\begin{eqnarray*}
D &=&\left( -1\right) ^{p_{1}}A\exp \left[ is\left( \Upsilon _{R}-\Upsilon
_{L}\right) /2+is\left( \xi _{R}+\xi _{L}\right) /2\right] , \\
B &=&\left( -1\right) ^{p_{2}}C\exp \left[ is\left( \Upsilon _{R}-\Upsilon
_{L}\right) /2-is\left( \xi _{R}+\xi _{L}\right) /2\right] .
\end{eqnarray*}
and the corresponding wavevectors are
\begin{eqnarray*}
q^{\left( \Sigma =+s\right) } &=&
-\frac{s\left( \Upsilon _{R}+\Upsilon _{L}\right) }{2L}-
\frac{\left( \xi _{R}-\xi _{L}\right) }{2L}+\frac{\pi p_{1}}{L}, \\
q^{\left( \Sigma =-s\right) } &=&
-\frac{s\left( \Upsilon _{R}+\Upsilon _{L}\right) }{2L}+
\frac{\left( \xi _{R}-\xi _{L}\right) }{2L}+\frac{\pi p_{2}}{L},
\end{eqnarray*}
where $\left\{ p_{1},p_{2}\right\} $ are integers such that $q\geq 0$.
The angles $\xi _{R}$ and $\xi _{L}$ break degeneracy whereas $\Upsilon_{R}$
and $\Upsilon _{L}$ break electron-hole symmetry. 

%%%%%%%%%%%%%%%%%%%%%%%%%%%%%%%%%%%%%%%%%%%%%%%%%%%%%%%%
\subsection{M3: mixed boundary conditions (i)}
%%%%%%%%%%%%%%%%%%%%%%%%%%%%%%%%%%%%%%%%%%%%%%%%%%%%%%%%

The boundary condition $\Psi =M_{3}\Psi$ produces the following relations
between the components of the wavefunction at the interface:
\begin{eqnarray*}
&&\psi _{AK}\left( 1-\sin \Omega _{u}\cos \zeta _{u}\right) +u\cos \Omega
_{u}e^{+i\eta -i\xi _{u}}\psi _{A\widetilde{K}} \\
&& \quad -u\sin \Omega _{u}\sin \zeta
_{u}e^{+i\eta }\psi _{BK} =0, \\
&&\psi _{BK}\left( 1+\sin \Omega _{u}\cos \zeta _{u}\right) -u\cos \Omega
_{u}e^{-i\eta -i\xi _{u}}\psi _{B\widetilde{K}} \\
&& \quad -u\sin \Omega _{u}\sin \zeta
_{u}e^{-i\eta }\psi _{AK} =0.
\end{eqnarray*}
The matrix $M_{3}$ has both diagonal and off-diagonal in K-point space parts.
Standing waves are created from linear combinations of all $\Psi
_{K}^{(R)}$, $\Psi _{\widetilde{K}}^{(L)}$, $\Psi _{\widetilde{K}}^{(R)}$
and $\Psi _{K}^{(L)}$.
We find that
\begin{eqnarray*}
B &=&\sin \Omega _{u}e^{isu\zeta _{u}}Ae^{iuqL}+isu\cos \Omega _{u}e^{-i\xi
_{u}}Ce^{iuqL}, \\
D &=&\sin \Omega _{u}e^{-isu\zeta _{u}}Ce^{iuqL}+isu\cos \Omega _{u}e^{+i\xi
_{u}}Ae^{iuqL}.
\end{eqnarray*}
and the corresponding wavevectors are given by
\begin{eqnarray}
&&\cos \left( 2qL\right) =\cos \beta ; \nonumber \\
&&\cos \beta=\sin \Omega _{R}\sin \Omega
_{L}\cos \left( \zeta _{R}+\zeta _{L}\right) \nonumber\\
&& \qquad - \cos \Omega _{R}\cos \Omega
_{L}\cos \left( \xi _{R}-\xi _{L}\right) , \nonumber
\end{eqnarray}
where $q \geq 0$ and $0 \leq \beta \leq \pi$.
The energy levels are
\begin{eqnarray}
E=sv\left\{ \frac{\beta}{2L},\frac{\pi n}{L}\pm \frac{\beta}{2L}
\right\} ,
\end{eqnarray}
where $n=\left\{ 1,2,3,\ldots \right\} $ are integers. The spectrum always
has positive-negative energy symmetry, but broken degeneracy for
$\beta \neq \{ 0,\pi\}$.

%%%%%%%%%%%%%%%%%%%%%%%%%%%%%%%%%%%%%%%%%%%%%%%%%%%%%%%%
\subsection{M4: mixed boundary conditions (ii)}
%%%%%%%%%%%%%%%%%%%%%%%%%%%%%%%%%%%%%%%%%%%%%%%%%%%%%%%%

The boundary condition $\Psi =M_{4}\Psi$ produces the following relations
between the components of the wavefunction at the interface: 
\begin{eqnarray*}
&&\psi _{AK}\left( 1+\sin \Theta _{u}\sin \zeta _{u}\right) -u\sin \Theta
_{u}\cos \zeta _{u}e^{+i\eta }\psi _{BK} \\
&& \quad -\cos \Theta _{u}e^{-i\xi _{u}}\psi
_{B\widetilde{K}} =0, \\
&&\psi _{BK}\left( 1-\sin \Theta _{u}\sin \zeta _{u}\right) -u\sin \Theta
_{u}\cos \zeta _{u}e^{-i\eta }\psi _{AK} \\
&& \quad -\cos \Theta _{u}e^{-i\xi _{u}}\psi
_{A\widetilde{K}} =0.
\end{eqnarray*}
The matrix $M_{4}$ has both diagonal and off-diagonal in K space
parts. Standing waves are created from linear combinations of all $\Psi
_{K}^{(R)}$, $\Psi _{\widetilde{K}}^{(L)}$, $\Psi _{\widetilde{K}}^{(R)}$
and $\Psi _{K}^{(L)}$. We find that 
\begin{eqnarray*}
B &=&isu\sin \Theta _{u}e^{isu\zeta _{u}}Ae^{iuqL}+\cos \Theta _{u}e^{-i\xi
_{u}}Ce^{iuqL}, \\
D &=&isu\sin \Theta _{u}e^{-isu\zeta _{u}}Ce^{iuqL}+\cos \Theta _{u}e^{+i\xi
_{u}}Ae^{iuqL}.
\end{eqnarray*}
and the corresponding wavevectors are given by 
\begin{eqnarray}
&&\cos \left( 2qL\right) = \cos \kappa ; \nonumber \\
&&\cos \kappa=\cos \Theta _{R}\cos \Theta
_{L}\cos \left( \xi _{R}-\xi _{L}\right) \nonumber\\
&&\qquad -\sin \Theta _{R}\sin \Theta
_{L}\cos \left( \zeta _{R}+\zeta _{L}\right) ,  \nonumber
\end{eqnarray}
where $q \geq 0$ and $0 \leq \kappa \leq \pi$.
The energy levels are
\begin{eqnarray}
E=sv\left\{ \frac{\kappa}{2L},\frac{\pi n}{L}\pm \frac{\kappa}{2L}
\right\} , 
\end{eqnarray}
where $n=\left\{ 1,2,3,\ldots \right\} $ are integers. The spectrum always
has positive-negative energy symmetry, but broken degeneracy for
$\kappa \neq \{ 0,\pi\}$.

%%%%%%%%%%%%%%%%%%%%%%%%%%%%%%%%%%%%%%%%%%%%%%%%%%%%%%%%
\section{Discussion}
%%%%%%%%%%%%%%%%%%%%%%%%%%%%%%%%%%%%%%%%%%%%%%%%%%%%%%%%

In this paper, we considered the effective mass model of spinless electrons
in single wall carbon nanotubes that describes slowly varying spatial envelope
wavefunctions $\Psi \left( {\bf q}, {\bf r} \right)$ with small wavevectors
${\bf q}$ in the region of linear dispersion $E = \pm v |{\bf q}|$ near 
the K-points.
Taking into account the two inequivalent K-points,
the envelope wavefunctions $\Psi$ obey the Dirac equation for massless fermions,
written in terms of four component spinors, with the role of spin assumed by
the relative amplitude of the wave function on the sublattice atoms (``A''
and ``B'').
We found that energy independent hard wall boundary conditions for the Dirac
equation may be written as $\Psi = M \Psi$ where $M$ is an Hermitian,
unitary $4\times 4$ matrix $M^{2}=1$ with the additional constraint that it
anticommutes with the component of the current operator normal to the boundary.
All possible linear combinations of matrices $M$ obeying these
constraints were expressed in terms of a small number of arbitrary parameters,
mixing angles, that describe mixing between boundary conditions with different
discrete symmetries. Then, in order to illustrate how the presence of non-zero mixing
angles breaks K-point degeneracy and electron-hole symmetry, we evaluated the
resulting energy level spectra for non-interacting electrons in finite-length
metallic nanotubes and the corresponding standing wave envelope functions.

The intention of this paper was to classify all possible boundary conditions
for the spatially long-range envelope functions of a closed nanotube
with length much greater than its circumference $L \gg L_c$.
The analysis was restricted to energy independent boundary conditions, although
in principle they could be generalised by performing a gradient expansion.
We focused on armchair tubes and did not consider the possibility of edge
states that may exist in zigzag graphite edges \cite{kle94,fuj96,ryu02}.
Rather than modelling the microscopic details of a boundary, such as
shape or roughness, the nature of the boundary is characterised by mixing angles
that describe the degree of symmetry breaking.
In practice, the correct choice of a particular set of boundary conditions and values of
symmetry mixing angles needed to describe a given nanotube will depend on experimental
details and may not be known beforehand.
To illustrate this, we compare our results to the idealised microscopic models of a
a capped armchair nanotube and a model of a boundary obtained by setting the
wavefunction to zero along a straight line of atoms.
We find that the description of a capped nanotube considered in
Ref.~\cite{y+a01} corresponds to our off-diagonal boundary conditions M2,
$\Psi =M_{2}\Psi$, with $\Upsilon_{u} = 0$ or $\pi$ so that the component
$\psi _{AK}$ of the wavefunction is related to $\psi_{B\widetilde{K}}$ at the
boundary.
A model of a boundary obtained by setting the wavefunction to zero,
which is equivalent to a particle-in-a-box model \cite{jiang02}, also corresponds to
our off-diagonal boundary conditions M2, $\Psi =M_{2}\Psi$,
but with $\Upsilon_{u} = \pm \pi/2$ so that the component $\psi _{AK}$ of the
wavefunction is related to $\psi_{A\widetilde{K}}$ at the boundary.
Although the two models are described by a different mixing angle $\Upsilon$, they both
correspond to the off-diagonal boundary conditions, introduce inter-valley scattering
at the boundary and, in general, they break K-point degeneracy with the mixing angle
$\xi_{u}$ dependent on the length of the nanotube.

%%%%%%%%%%%%%%%%%%%%%%%%%%%%%%%%%%%%%%%%%%%%%%%%%%%%%%%%%%%%%%%%%%%%%%%
\acknowledgements
The authors thank J~T~Chalker and C~J~Lambert for discussions,
and EPSRC for financial support.
%%%%%%%%%%%%%%%%%%%%%%%%%%%%%%%%%%%%%%%%%%%%%%%%%%%%%%%%%%%%%%%%%%%%%%%

%%%%%%%%%%%%%%%%%%%%%%%%%%%%%%%%%%%%%%%%%%%%%%%%%%%%%%%%%%%%%%%%%%%%%%%
\appendix
\section*{Appendix}
\setcounter{section}{1}

In this appendix, we briefly describe the method of finding linear
combinations of matrices that satisfy the constraints on $M$
described by the boundary conditions, Eq.~(\ref{bcintro}).
Any $4 \times 4$ matrix may be written as a linear combination
of the matrices $I_{\Pi } \otimes I_{\sigma }$,
$I_{\Pi } \otimes \left( {\bf n . \sigma } \right)$,
$\left( {\bf \nu . \Pi } \right) \otimes I_{\sigma }$, and
$\left( {\bf \nu . \Pi } \right) \otimes \left( {\bf n . \sigma } \right)$,
where ${\bf n }$ and ${\bf \nu }$ are arbitrary three-dimensional
space-like vectors.
The first step is to find all the linear combinations that produce
the unit matrix $I_{\Pi } \otimes I_{\sigma }$ when squared:
\begin{eqnarray}
M_a &=& I_{\Pi } \otimes I_{\sigma } , \nonumber \\
M_b &=& \cos \theta \left( I_{\Pi } \otimes  {\bf n_1 . \sigma } \right)
+ \sin \theta \left( {\bf \nu_1 . \Pi } \otimes
{\bf n_2 . \sigma } \right) , \nonumber \\
M_c &=& \cos \phi \left( {\bf \nu_1 . \Pi } \otimes I_{\sigma } \right)
+ \sin \phi \left( {\bf \nu_2 . \Pi } \otimes
{\bf n_2 . \sigma } \right) , \nonumber 
\end{eqnarray}
where the vectors are unit vectors with additional constraints
${\bf n_1 . n_2} = 0$, ${\bf \nu_1 . \nu_2} = 0$ that ensure no cross-terms
survive.
The next step is to find the conditions under which the matrices $M_a$,
$M_b$, and $M_c$ anticommute with the operator
${\bf n}_{{\bf B}}{\bf .\alpha }$ that is proportional to the component
of the current operator normal to the interface.
Clearly, $M_a$ does not anticommute, so it is
discarded.
The matrix $M_b$ anticommutes if
${\bf n_1 . n_B} = 0$, ${\bf n_2 . n_B} = 0$, and $\nu_1 = {\bf {\hat k}}$
(M1), or if
${\bf n_1 . n_B} = 0$, ${\bf n_2} = {\bf n_B}$, and $\nu_1$ is confined
to the $(x,y)$ plane (M3).
The matrix $M_c$ anticommutes if $\nu_1$ and $\nu_2$ are confined to the
$(x,y)$ plane and ${\bf n_2} = {\bf n_B}$ (M2), or if
$\nu_1$ is confined to the $(x,y)$ plane, $\nu_2 = {\bf {\hat k}}$,
and ${\bf n_2 . n_B} = 0$ (M4).

%%%%%%%%%%%%%%%%%%%%%%%%%%%%%%%%%%%%%%%%%%%%%%%%%%%%%%%%%%%%%%%%%%%%%%%

\end{document}